\documentclass[preprint2]{aastex63}

\usepackage{graphbox}
\usepackage[caption=false]{subfig}
\usepackage{CJK}

\accepted{March 15, 2021}
\submitjournal{ApJL}

\shorttitle{Predicting Solar EUV using Deep Learning}
\shortauthors{Pineci et al.}

\begin{document}

\begin{CJK*}{UTF8}{gbsn}

\title{Proxy-Based Prediction of Solar Extreme Ultraviolet Emission using Deep Learning}

\correspondingauthor{Peter Sadowski}
\email{peter.sadowski@hawaii.edu}

\author[0000-0002-6053-4974]{Anthony L. Pineci}
\affiliation{California Institute of Technology \\
Pasadena, CA 91125 USA}

\author[0000-0002-7354-5461]{Peter Sadowski}
\affiliation{Department of Information and Computer Sciences, University of Hawai'i at M\"{a}noa \\
1680 East-West Road \\
Honolulu, HI 96822 USA}


\author[0000-0002-5258-6846]{Eric Gaidos}
\affiliation{Department of Earth Sciences, University of Hawai'i at M\"{a}noa \\
1680 East-West Road \\
Honolulu, HI 96822 USA}

\author[0000-0003-4043-616X]{Xudong Sun (孙旭东)}
\affiliation{Institute for Astronomy, University of Hawai'i at M\"{a}noa \\
34 Ohia Ku Street\\
Pukalani, HI 96768 USA}



\begin{abstract}
High-energy radiation from the Sun governs the behavior of Earth's upper atmosphere and such radiation from any planet-hosting star can drive the long-term evolution of a planetary atmosphere.  However, much of this radiation is unobservable because of absorption by Earth's atmosphere and the interstellar medium. This motivates the identification of a proxy that can be readily observed from the ground.  Here, we evaluate absorption in the near-infrared 1083 nm triplet line of neutral orthohelium as a proxy for extreme ultraviolet (EUV) emission in the 30.4 nm line of \ion{He}{2} and 17.1 nm line of \ion{Fe}{9} from the Sun.  We apply deep learning to model the non-linear relationships, training and validating the model on historical, contemporaneous images of the solar disk acquired in the triplet \ion{He}{1} line by the ground-based SOLIS observatory and in the EUV by the NASA \emph{Solar Dynamics Observatory}.  The model is a fully-convolutional neural network (FCN) that incorporates spatial information and accounts for the projection of the spherical Sun to 2-d images. Using normalized target values, results indicate a median pixel-wise relative error of 20\% and a mean disk-integrated flux error of 7\% on a held-out test set. Qualitatively, the model learns the complex spatial correlations between \ion{He}{1} absorption and EUV emission has a predictive ability superior to that of a pixel-by-pixel model; it can also distinguish active regions from high-absorption filaments that do not result in EUV emission.
\end{abstract}



\section{Introduction} \label{sec:intro}

The Sun emits a mere $\sim10^{-6}$ of its energy in the extreme ultraviolet (EUV, $\lambda$=10-120 nm) but this radiation heats Earth's upper atmosphere, causing it to expand, and is a critical input for predictions of the lifetime of low-Earth-orbit satellites \citep{2018SpWea..16....5V}.  Models predict that over timescales of 0.1-1 Gyr, EUV radiation from host stars can drive significant escape of the atmospheres  of planets on close-in orbits \citep{2019AREPS..47...67O}, and the evolution in EUV emission as stars spin down and become less magnetically active is an area of active research in stellar and planetary astronomy \citep[e.g.,][]{Linsky2014}.  Problematically, the EUV is only observable from space and is heavily absorbed by the interstellar medium; only the Sun and some of the nearest normal stars have been detected.  Any proxy that can be monitored from the ground could greatly improve our understanding of EUV radiation from other stars.

Helium, an abundant element in all stars, has a neutral ``ortho" state with a triplet of absorption lines at 1083~nm (near-infrared) that is readily observed from the ground.  In cool stars, the metastable orthohelium state is primarily populated by recombination of singly-ionized He, which under quiescent (non-flaring) conditions is a product of photoionization by EUV ($\lambda < 50.4$ nm) photons \citep{2018ApJ...855L..11O}.  Neutral orthohelium is depleted both by ionizing UV photons with $\lambda < 259$ nm and de-excitation to the singlet state by electron collisions.  Thus there is a causal connection between EUV emission, which emanates from the transition region and corona \citep{Golding2017}, and the strength of \ion{He}{1} 1083 nm absorption, which arises in the upper chromosphere.  Three-dimensional radiation-magnetohydrodynamic simulations by \citet{Leenaarts2016} suggest that the He-ionizing radiation field has a diffuse, highly-scattered component from the corona and a localized component from the transition region.  The latter gives rise to spatial covariance between \ion{He}{1} absorption and EUV emission; both are elevated in active plage regions and co-vary  with time between solar minima and maxima \citep{Floyd2005}.  Long-term monitoring has established that the \ion{He}{1} 1083~nm line strength is an accurate proxy for EUV emission \citep{1994IAUS..154...59H,Deland2008}.  Since the advent of large-format near-infrared imaging arrays, the disk-resolved \ion{He}{1} line has been routinely monitored on the Sun \citep{Penn2014}.   The disk-integrated line has also been surveyed among mainly Sun-like stars, including planet hosts \citep{Andretta2017}.

\begin{figure*}[ht!]
\centering
\gridline{
\fig{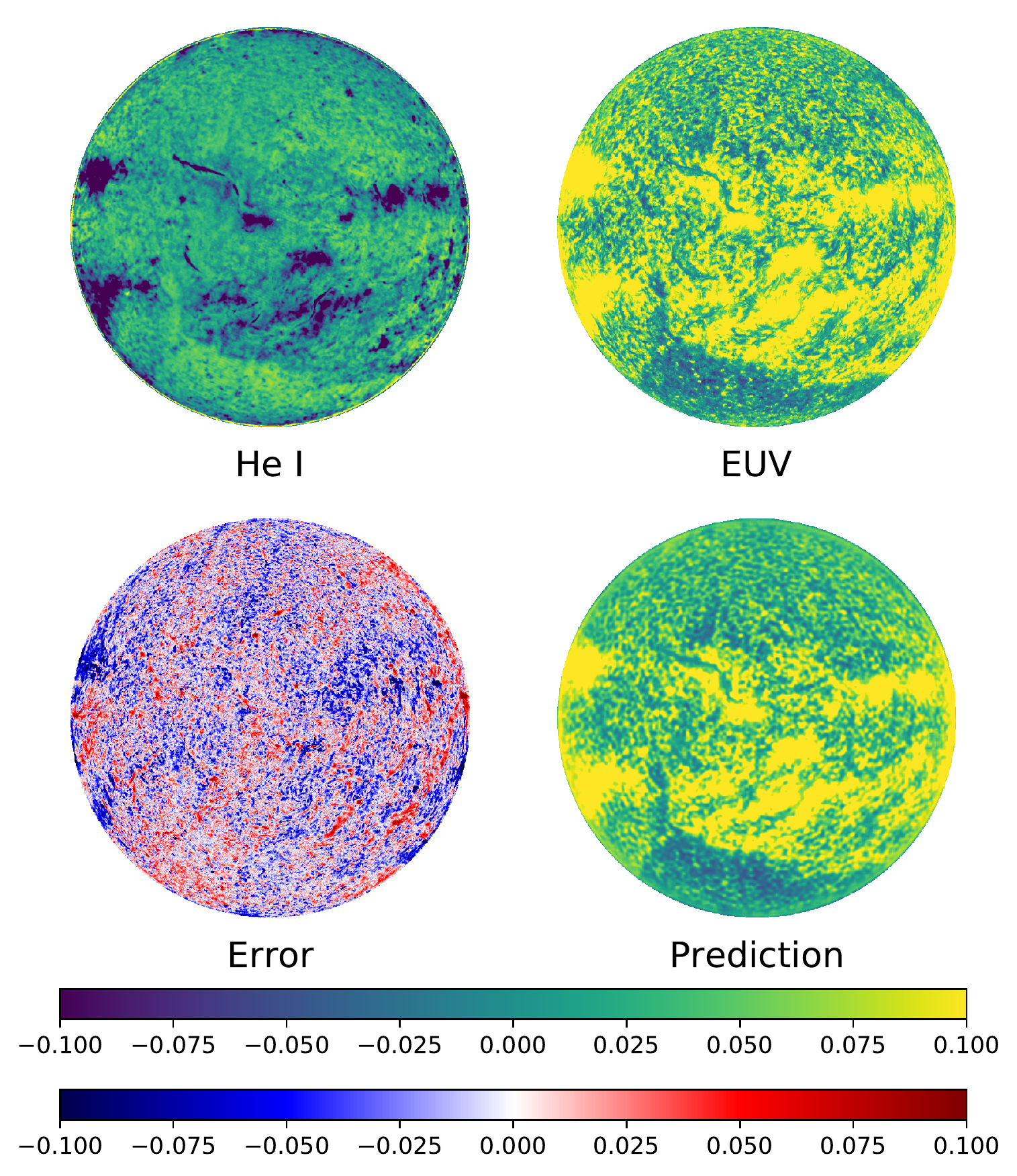}{0.48\textwidth}{(a)} \label{fig:pipeline}
\fig{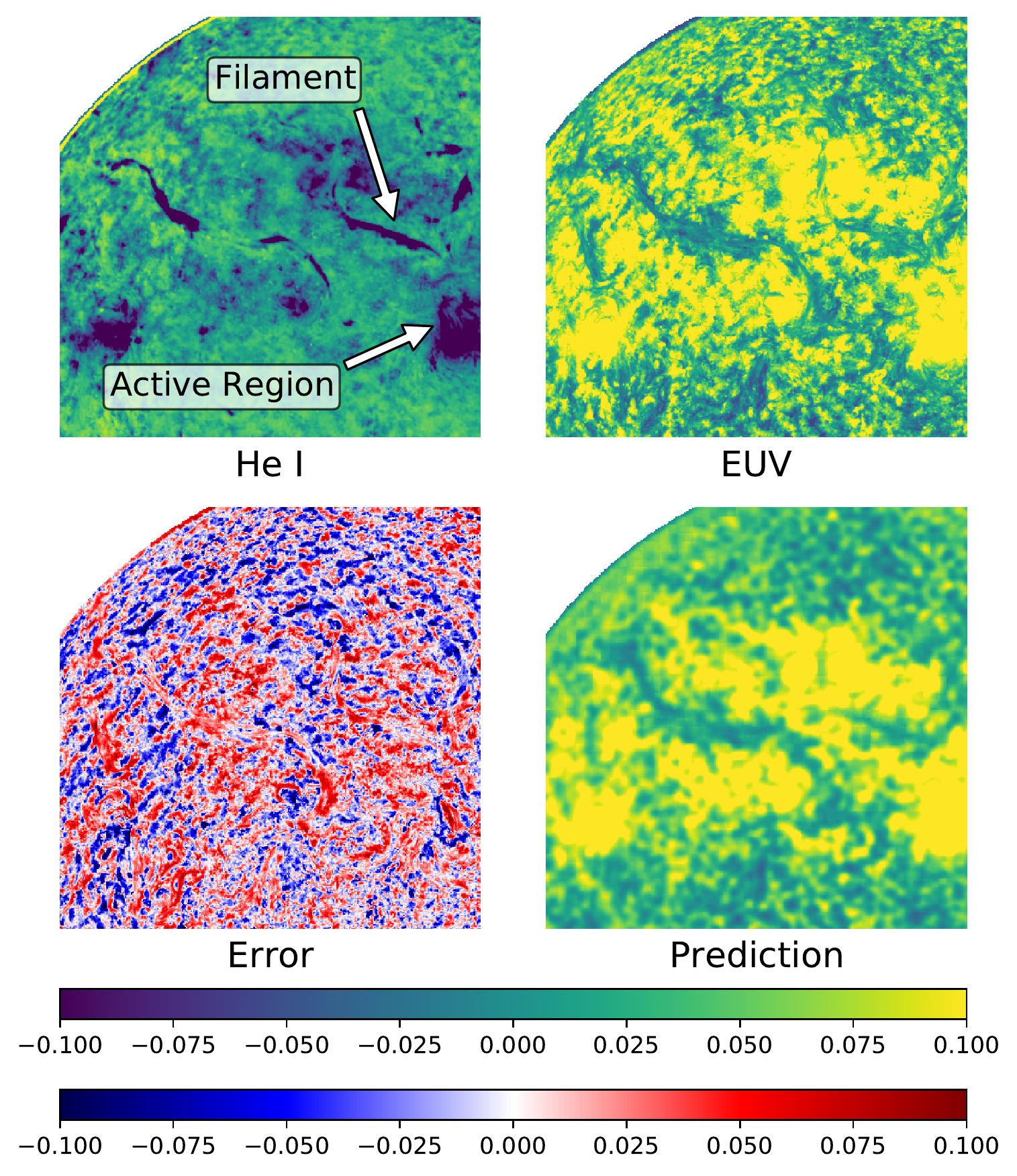}{0.48\textwidth}{(b)} \label{filaments}
}
\caption{(a) Deep neural networks are trained to map \ion{He}{1} absorption in the near-infrared at 1083 nm (top left) to the EUV emission of \ion{He}{2} at 30.4 nm (top right). Predictions on a test set (bottom right) correctly differentiate structures on the solar disk, and can be used to estimate total EUV emission with small and uniform error (bottom left). Units are standard deviation from the mean of the logarithmically-scaled values. (b) Close-up of a different region, showing that while \ion{He}{1} absorption and EUV emission are correlated, an exception is filaments, which are not associated with local sources of EUV emission. }
\end{figure*}

Due to the complexity of the solar atmosphere, particularly in magnetically active regions in which much of the EUV emission occurs, \emph{quantitative} prediction of the EUV-\ion{He}{1} relation is at the frontier of model physics and computational resources  \citep{Leenaarts2016,Rempel2017}. Given sufficient quantity and quality of data, empirical deep learning (DL) approaches can succeed where theoretical approaches cannot.  Here, we apply DL to map ground-based images of the 1083 nm \ion{He}{1} line strength to contemporaneous solar images of EUV emission obtained with the Solar Dynamics Observatory (SDO) space telescope on its geosynchronous orbit \citep{pesnell2012}.  The high-cadence, high spatial resolution, multi-wavelength aspects of SDO lends itself to DL \citep{Galvez2019}, and DL has been used to translate images of solar \ion{Ca}{2} emission into magnetic field maps (magnetograms) \citep{Shin2020}, and magnetograms into solar UV/EUV \citep{Park2019}.  DL has also been used on solar EUV images to predict coronal holes \citep{Illarionov2020} and solar wind intensity \citep{Upendran2020}.  

In this approach, a DL model is trained to accurately map infrared input (\ion{He}{1} 1083 nm) images to EUV output (\ion{He}{2} 30.4 nm) images.  The model's (hyper)parameters are tuned while evaluating the performance on a second, independent \emph{validation} set of infrared-EUV image pairs.  Finally, the ability of the model to predict EUV emission is then evaluated on a third, independent \emph{test} or \emph{``holdout"} dataset.  Our goals were to (i) identify the DL model that most accurately predict the disk-integrated EUV emission; and (ii) better understand the \emph{spatially local} relationship between \ion{He}{1} line strength and EUV emission, and any connection to specific kinds of structures in the stellar atmosphere.  A better understanding of this relationship could eventually lead to proxy-based estimates for replacing missing solar EUV data, and to estimates of EUV emission from distant planet host stars where direct measurements are presently not possible. 

\section{Datasets and Methods} \label{sec:methods}

We used full-disk solar images (i.e., line-scanning-based maps) of the \ion{He}{1} 1083~nm line strength (i.e., equivalent width) obtained by the Vector SpectroMagnetograph (VSM) at the Synoptic Optical Long-term Investigations of the Sun (SOLIS) observatory on Kitt Peak \citep{2009ASPC..405...47H}  (see Fig.~\ref{fig:pipeline}).  We paired these with full-disk images in the EUV obtained by the Atmospheric Imaging Assembly \citep[AIA;][]{2012SoPh..275...17L} instrument on SDO.  We primarily analyzed AIA images of emission in the \ion{He}{2} line at 30.4\,nm, the dominant line in the EUV spectrum of the quiet Sun \citep{2015A&A...581A..25D}.  Helium is very abundant (8.7\% by atom) and uniformly-distributed in the Sun and correlations between line of \ion{He}{1} and \ion{He}{2} are not due to any variation in abundance.

SOLIS operated from 2005 to 2015, whereas SDO has been operational since 2010.   A subset of 1321 image pairs was selected from data obtained during the overlap interval between May 2010 and July 2015. VSM images were obtained on a daily basis, weather permitting, while SDO images are obtained with a 12-second cadence, and are available for download at 2-minute cadence; therefore each \ion{He}{1} image was matched with the closest AIA image in time.  The offset in time between the \ion{He}{1} and EUV images in a pair is always less than 30 minutes and typically less than a minute.

AIA images are acquired at a plate scale of 0.6 arc-second, but these were spatially binned to a scale of 2.4 arc-second per pixel, then cropped to $864 \times 864$ pixels to remove space outside of the limb.  VSM $2048 \times 2048$ images with an original plate scale of 1 arc-second were resized using linear interpolation to match the size and plate scale of the cropped EUV images.  101 image pairs have missing or corrupted data due to instrument errors. Of these pairs, 68 have faulty AIA images, while the remainder have problematic SOLIS images.  AIA images are automatically metered according to the total solar intensity, and 28 images were obtained during solar flares and thus have short integration times and low signal-to-noise in regions of the solar disk outside of the flares.  These images were excluded to reduce systematic effects on the training. AIA are normalized according to their exposure time. The final processed dataset consists of 1221 matched pairs of \ion{He}{1} absorption and EUV emission. All image preprocessing is done using \textsc{astropy} \citep{astropy:2013, astropy:2018}, including using FITS headers for extraction of image exposure times, removing uncorrupted images, and matching the sun scale of input and output images.

AIA images are also affected by long-term degradation in detector sensitivity.  We corrected AIA pixel values using normalized values of sensitivity (telescope effective area) of the telescopes obtained from analysis of sounding-rocket calibrations \citep{2012SoPh..275...41B}, and linearly interpolating between calibration points in time.

EUV emission values have very large variance, so logarithmic values are used instead for DL training.  AIA data have been calibrated using electron hits on portions of the sensor \citep{Boerner2012}, a standardization which can result in negative emission values, so a constant term was added to the entire dataset before the log-transform so that $\sim1$ in 1 million values are clipped.

The performance of three neural network models was compared.  The primary model was a Fully Convolutional Network \citep[FCN,][]{long2015fully} based on a VGG16 architecture \citep{simonyan2014deep}, with $512\times512$ pixel single-channel input and output, and ``skip" connections (which bypass successive layers) with 8$\times$ up-sampling using transposed convolutions.  We also used variations on a pixel-wise predictor that contained no information from neighboring pixels (essentially an FCN with $1 \times 1$ kernels), and a U-Net architecture \citep{ronneberger2015u} that is capable of capturing features on the scale of the solar disk.

We account for some physical effects by including two additional input channels (features). The first is the Euclidean distance from the center of the disk to the limb.  This addresses the geometric projection of the spherical Sun onto a two-dimensional image and increasing distortion towards the limb, as well as enhanced absorption of EUV photons along the line of sight through the solar atmosphere.  The second additional channel is the solar latitude, which accounts for variation in the intensity and geometry of the solar magnetic field, the distribution and behavior of the active regions (plage and sunspots), and thus the spatial patterning of both \ion{He}{1} absorption and EUV emission.  North-south symmetry is assumed.  Any effects of the $\pm$7 degree annual variation in projection due to the solar obliquity were ignored.   

The data was split in order to minimize correlations between train and test images. The 1211 image pairs were grouped by month, then months were randomly assigned to training, validation, and test sets with proportions of 60\%, 20\%, and 20\%, respectively.

Models were implemented in \textsc{pytorch} \citep{paszke2019pytorch} and trained using the Adam optimization algorithm \citep{kingma2015adam}. 
The following hyperparameters were optimized for the FCN model by performing a grid search with \textsc{sherpa} \citep{hertel2020sherpa}: the learning rate, mini-batch size, dropout regularization, and the choice of training objective function --- either mean absolute error (MAE) or mean squared error (MSE). The best model (lowest validation set MSE) minimized the MAE objective using a learning rate of $0.0001$, mini-batch size of 2, dropout rate of $0.1$ in all layers, and a weight decay coefficient of $1\times10^{-7}$. The learning rate was decreased by a factor of $0.1$ when no improvement in validation score occurred for $5$ epochs, and training was stopped when no improvement occurred over 15 epochs (iterations through the training dataset).

\section{Results \label{sec:results}}

The performance of the different models is compared in Table~\ref{tab:performance} in terms of the MAE, root mean squared error (RMSE), and median absolute relative error computed over all predicted pixels in the test set. 

The convolutional neural networks that include information from neighboring pixels (U-Net and FCN) perform better than the models that make predictions for each pixel independently.  This shows that spatial features in the \ion{He}{1} image contain important information, e.g., for discriminating between active regions and filaments. However, the fully-convolutional network architecture with limited connectivity outperforms the U-Net architecture that is optimized to detect features on the scale of the entire disk.  This outcome can be explained by the fact that the spatial features (e.g., active regions and filaments) are restricted to small portions of the solar disk, so the U-Net's ability to capture large-scale features leads to over-fitting rather than providing additional useful information. We expect this effect to diminish with a larger training data set.

\begin{figure}
\centering
\includegraphics[trim=50 0 50 10, clip,width=\columnwidth]{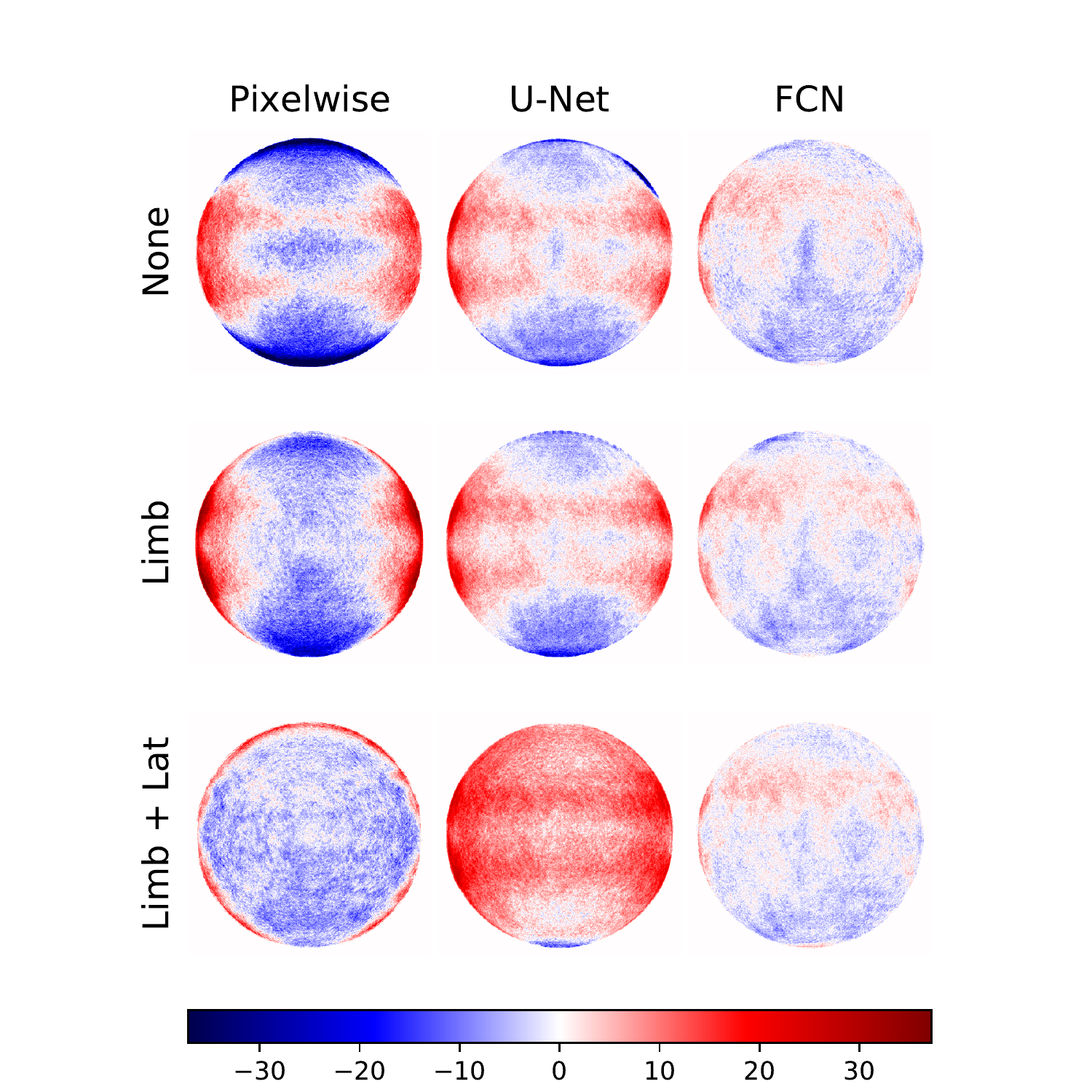}
\caption{Median of the signed relative percent error in pixel values using different neural network architectures.  Columns contain results from Pixel-wise, U-Net, and FCN models while rows contain results with the addition of physics-informed features. These images show how the error is correlated with latitude in models that do not include latitude as an input channel.}
\label{MedianRelativeError}
\end{figure}

Inclusion of limb distance and latitude as input features improves performance in the Pixel-wise and FCN models in terms of the training objective (not shown), and this translates to an improvement in the pixel-wise relative error (Fig. \ref{MedianRelativeError}). The same improvement is not seen with the U-Net model, which can be explained by the difference in spatial information available to the different architectures: the Pixel-wise and FCN models are restricted to using spatially-local information, while the U-Net model can use whole-disk information to make predictions. Thus, the addition of spatial features is less useful to the U-net model, and can even hurt performance by contributing to over-fitting.

\begin{deluxetable*}{l|rr|rr}
\tablecaption{Performance on test data. MAE and RMSE are given in terms of the units of the original data: counts/sec/pixel. \label{tab:performance}}
\tablehead{
\colhead{Model} & \colhead{MAE} & \colhead{RMSE} & \multicolumn2c{\% median/mean relative error}\\
& \colhead{pixel-wise} & \colhead{pixel-wise} & \colhead{pixel-wise} & \colhead{disk-integrated}
}
\startdata
Pixel-wise & $2.2\times10^{2}$ & $4.4\times10^{2}$  & 28.4 & 13.2\\
Pixel-wise + Limb & $2.1\times10^{2}$ & $4.2\times10^{2}$  & 27.1 & 12.8\\
Pixel-wise + Limb + Lat  & $2.0\times10^{2}$ & $7.8\times10^{2}$ & 27.0 & 11.6\\
\hline
U-Net  & $2.0\times10^{2}$ & $4.7\times10^{2}$ & 22.8 & 14.7\\
U-Net + Limb  & $2.0\times10^{2}$ & $4.7\times10^{2}$ & 22.7 & 15.2\\
U-Net + Limb + Lat  & $2.0\times10^{2}$ & $4.2\times10^{2}$ & 25.4 & 15.6\\
\hline
FCN  & $1.7\times10^{2}$ & $3.4\times10^{2}$ & 22.1 & 7.3\\
FCN + Limb  & $1.5\times10^{2}$ & $3.3\times10^{2}$ & 19.8 & 7.2\\
FCN + Limb + Lat & $1.5\times10^{2}$ & $3.3\times10^{2}$ & 19.9 & 7.0\\
\enddata
\end{deluxetable*}

\begin{figure}
\centering
\includegraphics[width=\columnwidth]{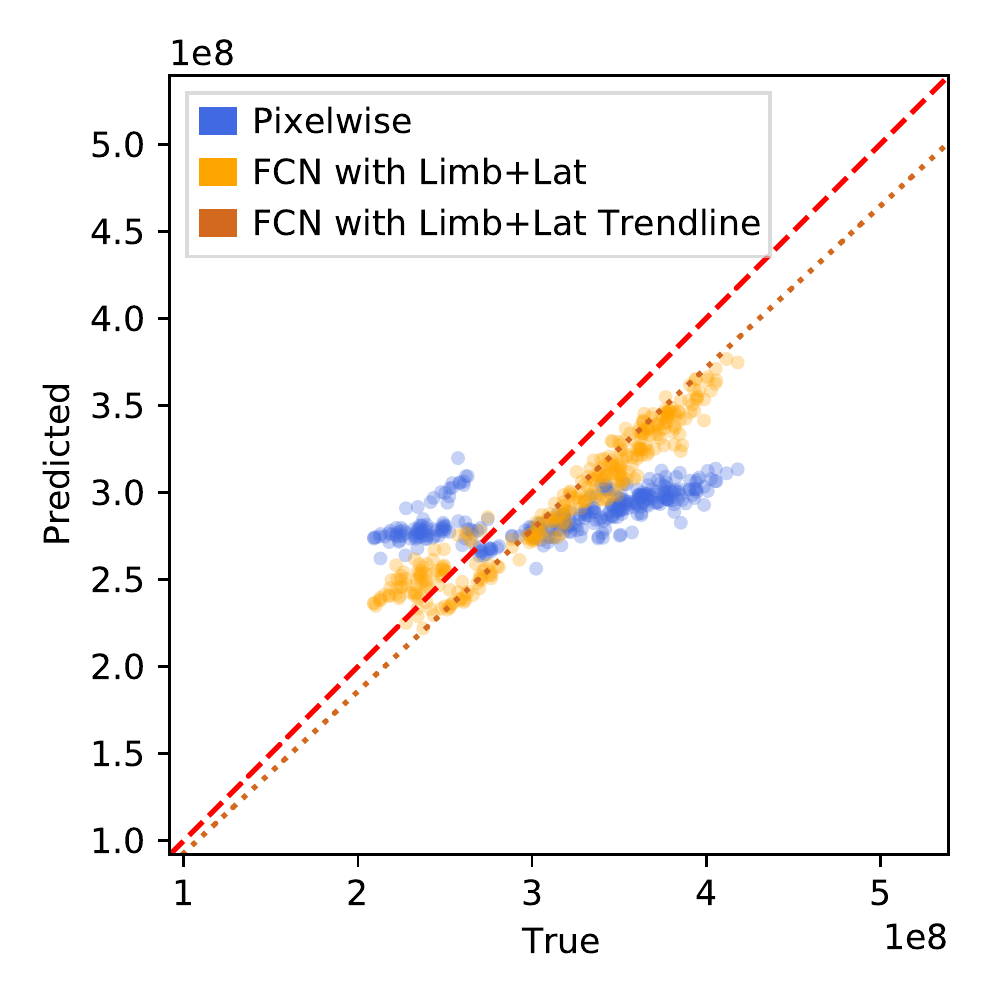}
\caption{Plot of predicted vs. observed disk-integrated flux for pixel-wise and FCN models on the test set. Perfect predictions would fall on the red dotted line, and the two models achieve a mean relative error of 13.2\% and 7.0\%, respectively. Training on log-scaled values results in a systematic underestimate of the flux; a linear regression model fit to the FCN predictions is shown in brown with slope 0.92.
}
\label{fig:flux}
\end{figure}

A prediction for disk-integrated flux in the 30.4 nm \ion{He}{2} line was obtained by summing the pixel values within the disk.  The best-performing FCN model also produces the most accurate prediction of the disk-integrated flux in terms of mean absolute relative error over the test images (Fig.~\ref{fig:flux}).

Qualitatively, the results indicate that the model learned some aspects of the spatial correlations between \ion{He}{1} absorption and EUV emission.  For example, an obvious exception to the trend for areas of high \ion{He}{1} absorption to correspond with EUV emission are filaments (aka prominences), thin, arcuate structures of magnetically-confined plasma suspended above the chromosphere \citep{2014LRSP...11....1P,2016A&A...589A..84K}. In these filaments, high \ion{He}{1} absorption corresponds to low EUV emission unlike other regions of the sun. The model is able to detect this behavior and correctly inverts the trend for absorption filaments (Fig. \ref{filaments}).

\section{Summary \& Discussion \label{sec:discussion}}

Using contemporaneous historical data from ground- and space-based solar telescopes for training data, we have demonstrated that a deep convolutional neural network can be used to predict the emission in a prominent EUV line (accessible only from space) from the absorption in an infrared line (accessible on the ground) with high accuracy.  We find that model performance can be improved through a physics-informed architecture design that (i) uses limited spatial information to discriminate between different physical phenomena including filaments and plage regions, and (ii) accounts for the projection of the Sun's surface onto a 2-d image.  

\begin{figure*}[ht!]
\centering
\subfloat[]{
\includegraphics[trim=0 0 0 0,clip,width=\columnwidth,align=t]{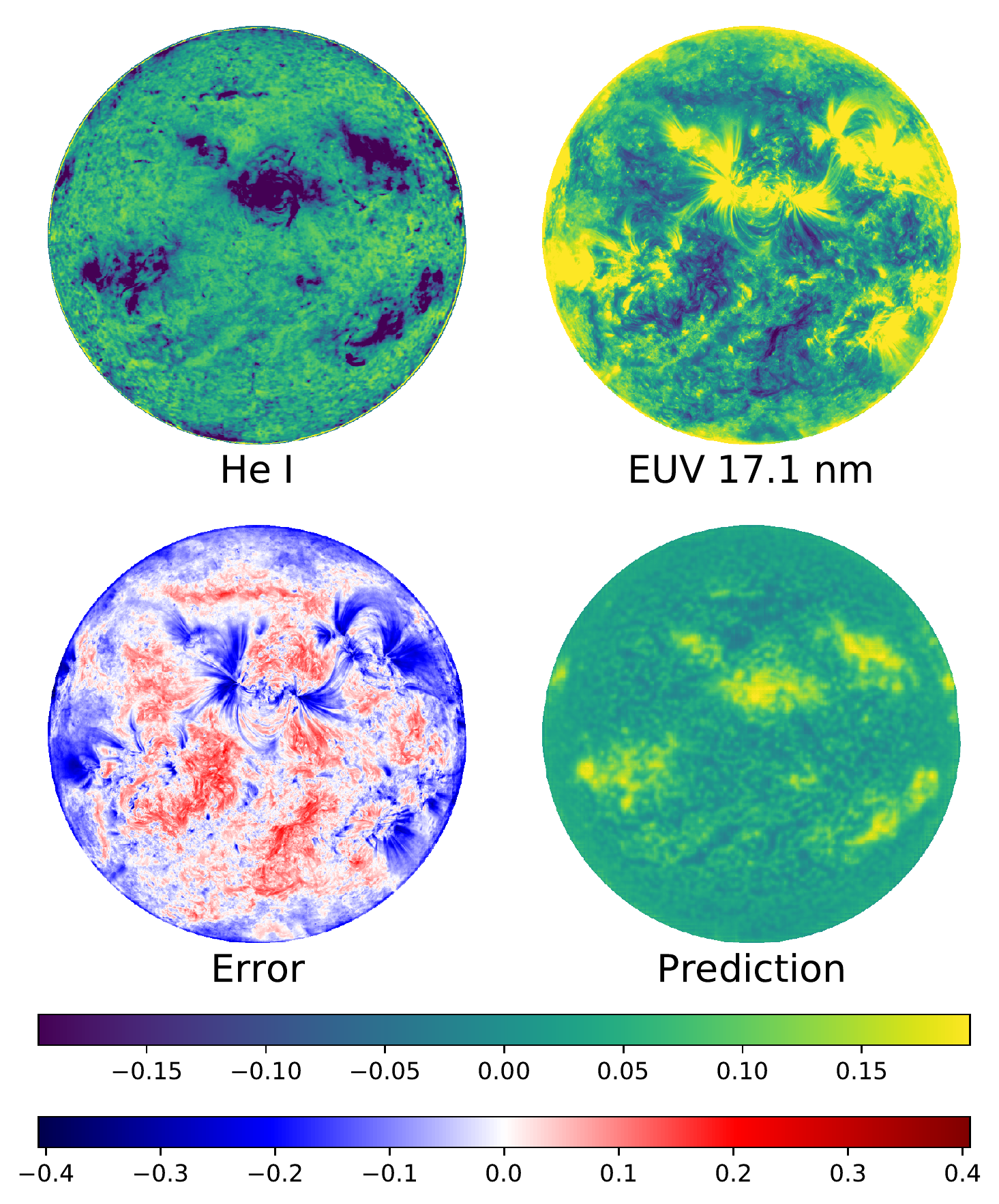}
}
\subfloat[]{
\includegraphics[trim=55 10 55 60,clip,width=\columnwidth,align=t]{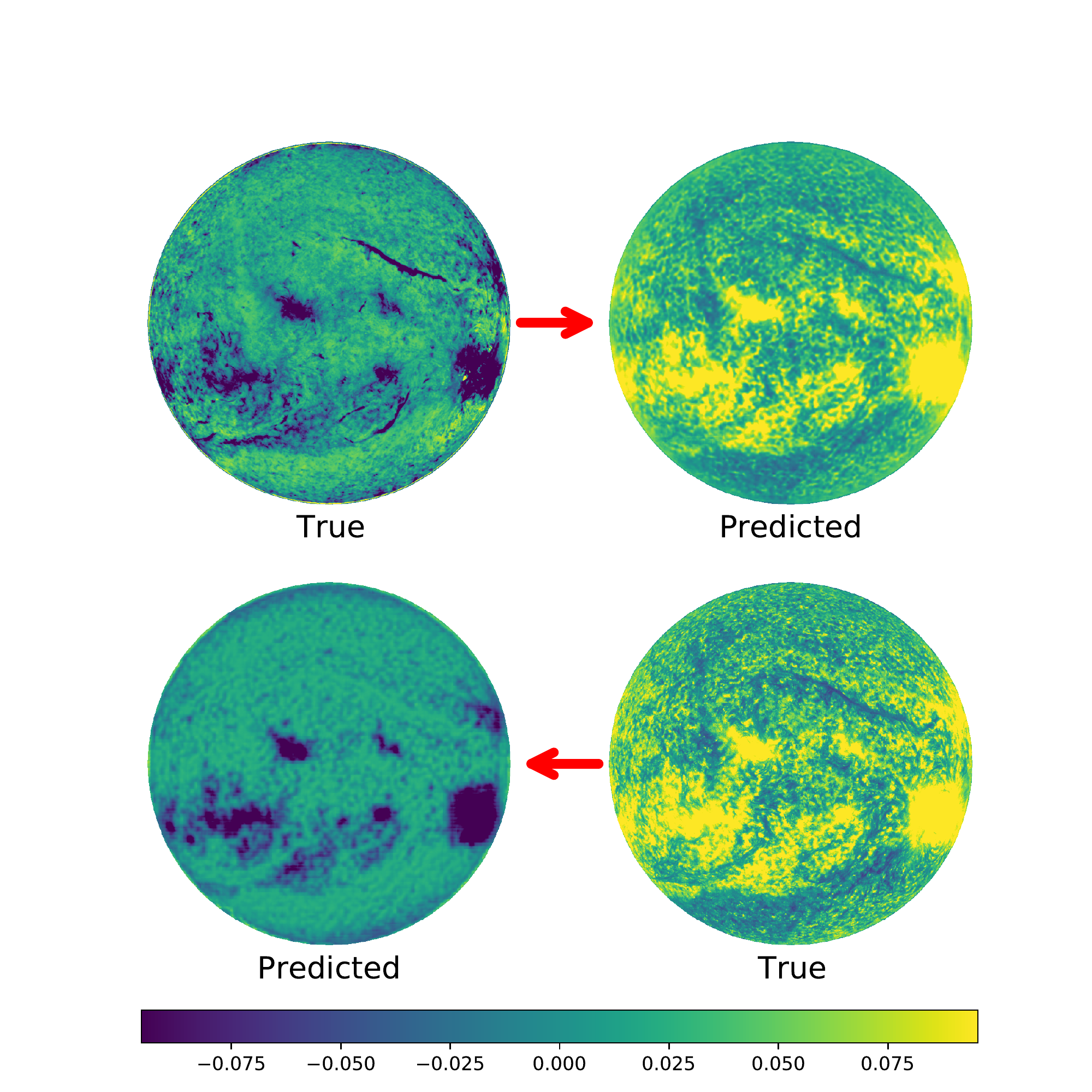}
}
\caption{(a) Contemporaneous solar images in \ion{He}{1} (top left) and \ion{Fe}{9} 17.1 nm  (top right). Predictions of 17.1 nm emission (bottom right) have higher error (bottom left) than predictions of emission at 30.4 nm. (b) Example of a SOLIS image of \ion{He}{1} 1083~nm absorption (top left) and the predicted AIA image of \ion{He}{2} 30.4\,nm  (top right) using the best-performing FCN model. The same architecture is then used to predict the \ion{He}{1} image (bottom left) from the observed EUV image (bottom right).}
\label{fig:171Predictions}
\end{figure*}

The SDO AIA images the Sun at nine other wavelengths between 9.4 and 450 nm, each of which probes different temperatures and regions of the solar atmosphere \citep{2012SoPh..275...17L}.  We used the FCN model to predict emission in another prominent EUV line, that of \ion{Fe}{9} at 17.1\,nm.  Figure \ref{fig:171Predictions}a shows a representative prediction and its errors on an out-of-sample image. The resulting performance of this model is larger than for the corresponding 30.4~nm model ($26.1\%$ vs. $19.9\%$ pixel-wise median relative error on the test set). This is expected because 17.1\,nm emission is probing a region that is higher in the solar atmosphere, its contribution to the formation of triplet He I at any given point will be more spatially dispersed, and thus finer scales in patterns of emission in the 17.1 nm line will not be captured.   Although predictions capture the overall distribution of emission (bottom right panel of Fig. \ref{fig:171Predictions}a), filamentous structures that reflect the magnetic field topology in that region of the atmosphere are not reproduced.

We also trained a FCN model in reverse, swapping the model inputs and outputs, to test whether aspects of physical causality are evident. Although it is for practicality that we predict EUV emission at 30.4 nm (the less accessible observation) based on \ion{He}{1} absorption at 1083~nm (the more accessible observation), it is EUV photoionization of He and subsequent recombination that produces triplet \ion{He}{1}.  Figure \ref{fig:171Predictions}b compares one representative prediction to the observation for each directions.  The pixel-wise median relative error of $19.9\%$ for \ion{He}{1} $\rightarrow$ EUV sense (top) increases to $24.1\%$ for EUV $\rightarrow$ \ion{He}{1} (bottom). One marked difference is the inability of the EUV $\rightarrow$ \ion{He}{1} to predict the existence of filaments.  Since the \ion{He}{1} absorption is not \emph{locally} driven it cannot be predicted by the \emph{local} distribution of EUV emission.

Our results suggest that at least partial reconstruction of EUV emission could be achieved using He\,I observations, either to reconstruct missing data at past epochs, or to estimate EUV emission at future time when EUV observations from space are interrupted or become unavailable.   The Extreme ultraviolet Imaging Telescope (EIT) on the Solar and Heliospheric Observatory (SOHO) was obtaining full disk images in four EUV lines, including that of \ion{He}{2} at 30.4 nm, for 15 years, albeit at much lower cadence and spatial resolution than the AIA \citep{Delaboudiniere1995}, and represents another potential opportunity to test the performance of DL models, particularly in the 5-year overlap interval with SOLIS.  Future work could also test whether the inclusion of other imaging data from ground-based observatories, i.e., H$\alpha$, \ion{Ca}{2}~HK, or magnetograms \citep[like that produced by SOLIS-VSM,][]{Gosain2013} can improve predictions.  Application to other stars is inhibited by the absence of resolve-disk information, but potentially such DL models could be used as computationally efficient means to convert multiple disk-integrated observables, including triplet \ion{He}{1} absorption at 1083 nm, into total EUV emission using descriptions of the distribution of activity with a few tunable parameters.  

\acknowledgements

Tom Schad offered valuable comments on a draft of this manuscript.  ALP was supported in part by Samuel P. and Frances Krown through the Caltech Summer Undergraduate Research Fellowship program.  This material is based upon work supported by the National Science Foundation under Grant No. 2008344. Advanced computing resources from the University of Hawai`i Information Technology Services Cyberinfrastructure are gratefully acknowledged.  EG acknowledges support as a long-term visitor in the Center for Space and Habitability at the University of Bern.

\bibliographystyle{aasjournal}
\bibliography{refs,nn}

\end{CJK*}

\end{document}